\documentclass[aps,twocolumn,amsmath,amssymb,showpacs,showkeys,superscriptaddress]{revtex4-1}

\usepackage{graphicx}

\begin{document}

\title{Numerical model for atomtronic circuit analysis}

\author{Weng W. Chow}
 \affiliation{Sandia National Laboratories, Albuquerque, New Mexico 87185-1086}

\author{Cameron J. E. Straatsma}
 \affiliation{JILA and Department of Electrical, Computer, and Energy Engineering, University of Colorado, Boulder, Colorado 80309-0440}
\author{Dana Z. Anderson}
 \email{dana@jila.colorado.edu}
 \affiliation{JILA and Department of Physics, University of Colorado and National Institute of Standards and Technology, Boulder, Colorado 80309-0440}

\date{\today}

\begin{abstract}
A model for studying atomtronic devices and circuits based on finite temperature Bose-condensed gases is presented. The approach involves numerically solving equations of motion for atomic populations and coherences, derived using the Bose-Hubbard Hamiltonian and the Heisenberg picture. The resulting cluster expansion is truncated at a level giving balance between physics rigor and numerical demand mitigation. This approach allows parametric studies involving time scales that cover both the rapid population dynamics relevant to non-equilibrium state evolution, as well as the much longer time durations typical for reaching steady-state device operation. The model is demonstrated by studying the evolution of a Bose-condensed gas in the presence of atom injection and extraction in a double-well potential. In this configuration phase-locking between condensates in each well of the potential is readily observed, and its influence on the evolution of the system is studied.
\end{abstract}

\maketitle

\section{Introduction}
In many respects atomtronic devices and circuits~\cite{article:JQI_hysteresis,article:atomtronics_Zozulya} can be understood using the same framework applicable to their electronic counterparts.  Analytical tools such as Kirchhoff's voltage and current laws, for example, are simply circuit-relevant statements about energy and particle conservation, which apply equally well to electronic and atomtronic systems. Operating in the ultracold regime, two aspects of atomtronic circuits make them both interesting and very different from classical electronics. First, quantum coherence can be long lived compared with other circuit timescales of interest.  Second, atomtronic circuits are typically decoupled from a thermal reservoir~\cite{article:atomtronics_Holland}. An active circuit is by definition a non-thermal-equilibrium dynamical system and any useful circuit will generate heat for both fundamental and technical reasons. In general, this means that the temperature will vary from place-to-place across the circuit.  Thermal coupling to the environment enables an electronic circuit to reach thermal steady-state (not thermal equilibrium) and finite device temperature.  However, heat has no thermal bath to flow to in an atomtronic circuit.  While doubling its temperature often leads to permanent failure of a nominal room-temperature electronic device, the same is not true of an atomtronic device.  Yet such significant changes in temperature can certainly have an impact on the quantum character of devices operating in the ultracold regime.  The interplay between the thermal-statistical and quantum aspects of atomtronic circuits at the physical level is profoundly related to circuit functionality through entropy -- physical entropy on the one hand and information entropy on the other.  For example, if a circuit function corresponds to the lowering of information entropy, as is often the case, this must correspond to a lowering of physical entropy in some region of the circuit, which demands that there is a corresponding increase in entropy in another region of the circuit if the second law of thermodynamics is to be upheld.

Electronic circuits are challenging from an analytical standpoint and atomtronic circuits are dramatically more so. Only the most elementary electronic circuits can be treated from first principles; complex circuits are designed using heuristic principles and analyzed using parameterized device models and highly optimized numerical solvers.  Our purpose in this work is to make some headway along these same lines in the atomtronics domain.  We motivate our objective with a few simple questions regarding a conceptually simple system. As it turns out, the questions are very difficult to answer.

\begin{figure}
\centering
\includegraphics[scale=1]{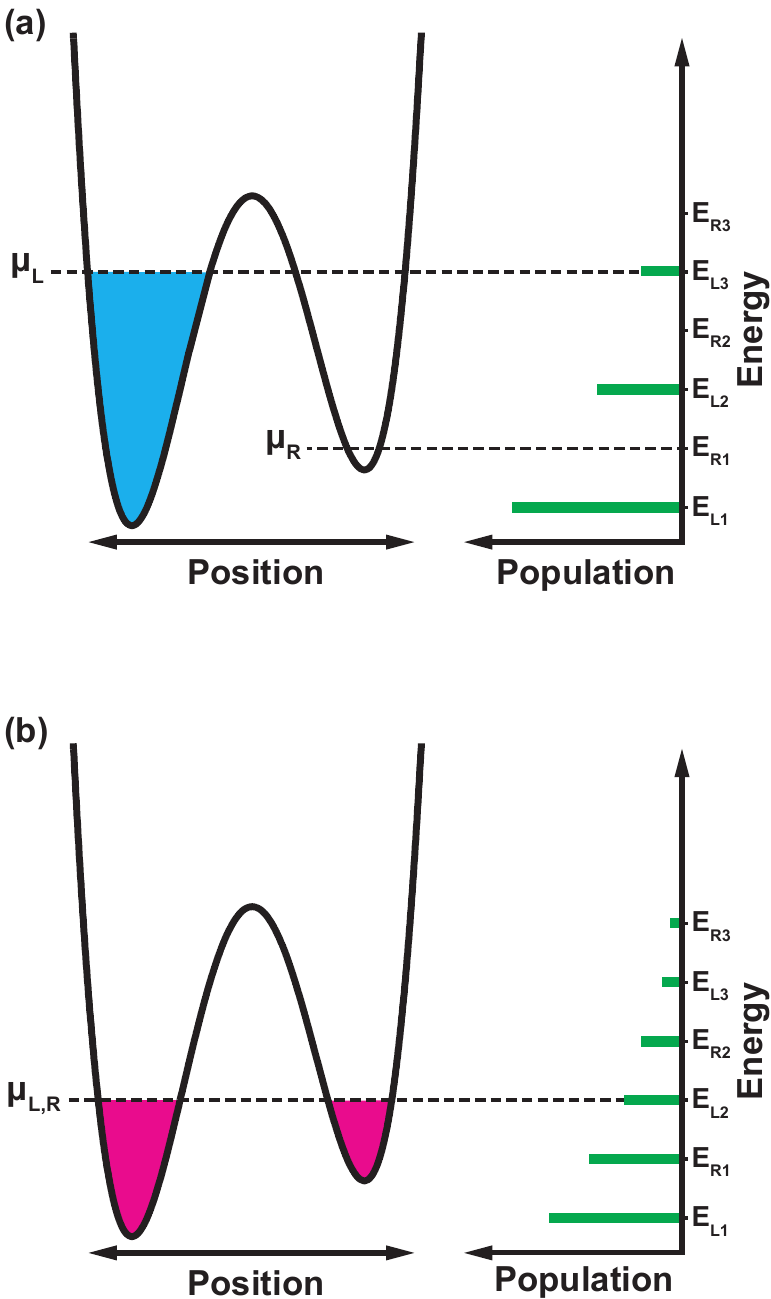}
\caption{\label{fig:fig1}Schematic representation of a BEC heating as it comes to thermal equilibrium in a double-well potential. (a) Initially, the state of the system is highly non-equilibrium as demonstrated by the empty states in the population distribution at energies corresponding to levels in the right-hand well. (b) In thermal equilibrium the population has been redistributed via elastic collisions between atoms because population of lower lying states in the right-hand well are always accompanied by population of higher lying states. The net result is heating of the initial distribution as depicted by the change in color from (a) (blue, colder) to (b) (red, hotter).}
\end{figure}

Consider the double-potential-well system of Fig.~\ref{fig:fig1}(a).
The right-hand well is initially devoid of atoms while the left-hand well holds a Bose-Einstein condensate (BEC) at, or very near to, absolute zero.  The BEC is characterized by a chemical potential, $\mu_{L}$, while the empty right-hand well has a chemical potential equal to its ground state energy.  For this discussion we take $\mu_{L}$ to be  large compared to the single-particle energy level spacing near the ground state of the well, but considerably less than the barrier height between the wells.  Our first question is: ``Does this (isolated) system come to thermal equilibrium, and, if so, by what route does it do so?"  By thermal equilibrium we mean the system is described by a single, time-independent temperature, $T$, and chemical potential, $\mu$, as illustrated in Fig.~\ref{fig:fig1}(b).  We note that the equilibrium state must have finite temperature to maintain fixed total internal energy, even if the initial left-hand BEC is at zero temperature.  Thus, by whatever route the system arrives at equilibrium, heating must take place.  The energy associated with the temperature rise comes at the expense of chemical potential, which necessarily decreases from its initial value in the left-hand well.  Supposing that parameters are chosen such that thermal equilibrium supports a condensate in both wells, then their chemical potentials must be equal in thermal equilibrium.  This implies that their respective wavefunctions have a well-defined relative phase which, ignoring fluctuations, is constant in time.  Our second question is: ``Can the two condensates phase-lock, and if so, under what conditions does this occur?"  By ``phase-lock" we mean that if the relative phase is perturbed it subsequently returns to a fixed equilibrium value.  Phase-locking is a ubiquitous phenomenon in nature (see~\cite{book:synchronization} for a number of examples), and more to our point, it is an important functional primitive among electronic circuits -- it lies at the heart of every modern radio receiver, for example. In the circuit context, questions regarding thermal equilibrium can similarly be asked regarding dynamical equilibrium.

To make an attempt at answering both questions in the circuit context, we developed a model of an atomtronic circuit operating with finite temperature BECs in the presence of atomic current flows, which are determined by atom injection and extraction processes, as well as collisions. The model draws upon a wealth of theoretical developments involving weakly interacting Bose gases~\cite{article:Jackson_tutorial,article:Blakie,article:nbody,article:Ilo-Okeke}. Past studies of phenomena occurring in multi-well potentials include atom interferometry~\cite{article:Ilo-Okeke}, the bosonic Josephson effect~\cite{article:Mazzarella}, dynamical phase-locking of BECs~\cite{article:Boukobza}, and transistor-like behavior of ultracold atoms~\cite{article:Stickney}. They typically involve solutions of the time-independent Schr\"{o}dinger equation, or the more general von Neumann equation for the Bose-Hubbard Hamiltonian~\cite{article:Gersch}, and treat collision effects at the level of the Boltzmann equation~\cite{article:nbody,book:Huang}. In these studies, detailed analyses were performed, sometimes under restrictive experimental conditions, such as for a microcanonical ensemble with a fixed total number of atoms to isolate the phenomenon of interest in order to understand its underlying physics. The methods are effective at extracting intrinsic and extrinsic parameters that characterize the gas, such as damping rates for various collective excitations. These parameters provide good anchors for our model, which is more phenomenologically based for two reasons. First, the number of atoms in an atomtronic circuit is not known \emph{a priori} and is likely not conserved. Therefore, one has to solve the more general time-dependent problem accounting for the influence of atom injection and extraction on BEC dynamics. Second, the necessary computations cannot be so demanding as to prevent timely and comprehensive parametric studies.

Our starting point remains the Bose-Hubbard Hamiltonian. However, we choose to work in the Heisenberg picture and use a cluster expansion approach to circumvent the complication of a rapidly increasing configuration space with an increasing number of atoms. The approach involves tracking the time evolution of atomic populations in the stationary states of the confining potential. Many-body equations of motion for atomic populations and coherences are derived. They contain hopping effects, energy renormalizations, and coherences within the context of a mean-field approximation. Collisions giving rise to relaxation and dephasing are modeled phenomenologically using an effective relaxation rate approximation, with the effective relaxation rates obtained from experiment or quantum kinetic calculations~\cite{article:nbody}, for example. This type of treatment is common in quantum electronics~\cite{book:laser_phys}. Beyond the investigation of phase-locking, our interest is to understand the extent to which the considerable body of work in many-body semiconductor laser theory~\cite{book:Koch_quantum,book:Koch_lasers} can provide tools for designing and analyzing atomtronic devices and circuits.

Within our model a trade-off is made between rigor and mitigation of numerical demands. The result is an approach allowing parametric studies involving timescales spanning over three orders of magnitude. With this dynamical range rapid population dynamics relevant to non-equilibrium effects can be modeled, as well as the long time durations typical of steady-state device operation. The tracking of slow device dynamics is extremely challenging for the rigorous quantum kinetic approaches. Lastly, limiting the computational demand allows other details of an atomtronic circuit to be included, such as those involving atom injection and extraction processes, as well as complex confinement potentials. The end product is a predictive and flexible model that can be used for designing atomtronic devices and analyzing experimental data.

\begin{figure}
\centering
\includegraphics[scale=1]{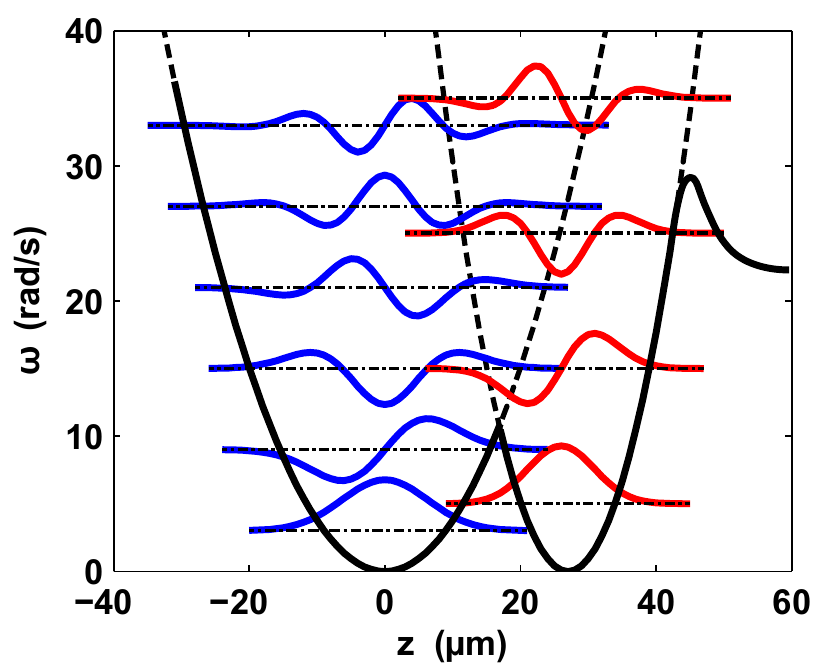}
\caption{\label{fig:fig2} Confining potential (solid, black) for the double-well geometry being studied. The parabolas (dashed, black) represent the potentials for the uncoupled wells. Also plotted are the eigenfunctions (solid, blue and red) of each well displaced vertically according to energy.}
\end{figure}

Using the model, we address BEC phase-locking in a system that is simplified, but also geared more towards circuitry than the fully isolated system of Fig.~\ref{fig:fig1}. Figure~\ref{fig:fig2} shows the model potential used for our current study.
The system consists of two, weakly coupled harmonic wells of different frequency where atoms are injected into the left-hand well and extracted from the right-hand well. Additionally, atoms are extracted from the highest energy states in each well analogous to the process of evaporative cooling used in typical BEC experiments. Notably, establishment of a phase relationship between coupled BECs has been studied previously in a similar geometry~\cite{article:Burnett_phase} where dissipation and atomic interactions were found to play a key role in the phase-locking mechanism. In this work we study this phase evolution under the influence of atom injection; and, as will be shown below, the strength of coupling between the ground states of each well strongly dictates the steady-state behaviour of the system.

The remainder of the paper is organized as follows. Section II provides a description of the theory and model development. The working many-body equations of motion are presented and the collision model described. Section III illustrates an application of the model to study atom transport between two adjacent harmonic potentials in the presence of atom injection and extraction. The simulations show two distinct dynamical scenarios involving the phase-locking of states in the different wells. Section IV summarizes the results and discusses future improvements to the model.

\section{Theory}
The system Hamiltonian can be written as the sum of two parts:
\begin{equation}
H=\sum_{i}\left[ \frac{p_{i}^{2}}{2m}+V\left(\textbf{r}_{i}\right) \right]+\sum_{i>j}V\left(\textbf{r}_{i},\textbf{r}_{j}\right),
\end{equation}
where $p_{i}$ is the momentum of the $i^{th}$ atom (all with mass $m$), $V\left(\textbf{r}_{i}\right)$ is the confining potential, and $V\left(\textbf{r}_{i},\textbf{r}_{j}\right)$ describes interatomic interactions, which play an important role in the population dynamics of the system and are often described by a contact interaction~\cite{article:Blakie,article:nbody}. Although the model readily describes an arbitrary number of wells, for brevity we present the derivation assuming a double-well potential. In this case, we introduce the Hamiltonians for the uncoupled well potentials, $V_a\left(r\right)$ and $V_b\left(r\right)$:
\begin{eqnarray}
H_{A} &=&\frac{p^{2}}{2m}+V_{a}\left(\textbf{r}\right), \nonumber \\
H_{B} &=&\frac{p^{2}}{2m}+V_{b}\left(\textbf{r}\right), \label{eqn:H0}
\end{eqnarray}
which have the solutions
\begin{eqnarray}
H_{A}\phi_{i}^{A}\left(\textbf{r}\right) & = & \epsilon_{i}^{A}\phi_{i}^{A}\left(\textbf{r}\right), \nonumber \\
H_{B}\phi_{i}^{B}\left(\textbf{r}\right) & = & \epsilon_{i}^{B}\phi_{i}^{B}\left(\textbf{r}\right). \label{eqn:eigdecomp}
\end{eqnarray}
With $V_{a}\left(\textbf{r}\right)$ and $V_{b}\left(\textbf{r}\right)$ given by the dashed parabolas depicted in Fig.~\ref{fig:fig2}, Eq.~\eqref{eqn:eigdecomp} describes the familiar eigenvalue equation of the quantum harmonic oscillator.

Using these eigenstates, we introduce the second-quantized wavefunction of the double-well system:
\begin{equation}
\hat{\psi}\left(\textbf{r}\right) = \sum_ia_i\phi_i^A\left(\textbf{r}\right)+\sum_jb_j\phi_j^B\left(\textbf{r}\right),
\end{equation}
where $\phi_i^A\left(\textbf{r}\right)$ and $\phi_j^B\left(\textbf{r}\right)$ are localized in the left- and right-hand wells, respectively, and $a_i$ and $b_j$ are the annihilation operators for atoms in the $i^{\text{th}}$ and $j^{\text{th}}$ state of the left- and right-hand wells, respectively. The creation and annihilation operators obey the commutation relations,
\begin{eqnarray}
\left[a_i,a_j\right] & = & \left[b_i,b_j\right] = 0, \\
\left[a_i,b_j\right] & = & \left[a_i,b_j^\dagger\right] = 0, \\
\left[a_i,a_j^\dagger\right] & = & \left[b_i,b_j^\dagger\right] = \delta_{i,j}.
\end{eqnarray}
Thus, in second-quantized form,
\begin{eqnarray}
&& H = \int d^{3}\textbf{r}~\hat{\psi}^{\dagger}\left(\textbf{r}\right)\left[\frac{p^{2}}{2m}+V\left(\textbf{r}\right)\right]\hat{\psi}\left(\textbf{r}\right) \nonumber \\
&& +\int d^{3}\textbf{r}^{\prime}\int d^{3}\textbf{r}~\hat{\psi}^{\dagger}\left(\textbf{r}^{\prime}\right)\hat{\psi}^{\dagger}\left(\textbf{r}\right)V\left(\textbf{r},\textbf{r}^{\prime }\right)\hat{\psi}\left(\textbf{r}\right)\hat{\psi}\left(\textbf{r}^{\prime}\right),
\end{eqnarray}
giving essentially the two-site Bose-Hubbard Hamiltonian~\cite{article:Gersch}:
\begin{eqnarray} \label{eqn:mbHamil}
H & = & \sum_i\epsilon_i^A a_i^\dagger a_i+\sum_j\epsilon_j^B b_j^\dagger b_j \nonumber \\
&& +\sum_{i,j}\left(g_{ij} a_i^\dagger b_j+H.a.\right) \nonumber \\
&& +\sum_{i,j,k\neq0}\left(v_{ijk}^{AA} a_{i-k}^\dagger a_{j+k}^\dagger a_j a_i+H.a.\right) \nonumber \\
&& +\sum_{i,j,k\neq0}\left(v_{ijk}^{AB} a_{i-k}^\dagger b_{j+k}^\dagger b_j a_i+H.a.\right) \nonumber \\
&& +\sum_{i,j,k\neq0}\left(v_{ijk}^{BB} b_{i-k}^\dagger b_{j+k}^\dagger b_j b_i+H.a.\right),
\end{eqnarray}
where $g_{ij}$ describes a direct coupling between the two wells and $v_{ijk}$ accounts for interatomic correlations. These two coefficients are given by
\begin{eqnarray}
g_{ij} & = & \int d^{3}\mathbf{r}~\phi_{i}^{A}\left(\textbf{r}\right)\left[\frac{p^{2}}{2m}+V\left(\textbf{r}\right)\right]\phi_{j}^{B}\left(\textbf{r}\right), \label{eqn:gij} \\
v_{ijk}^{\alpha,\beta} & = & \int d^{3}\textbf{r}^{\prime}\int d^{3}\textbf{r}~\phi_{i-k}^{\alpha}\left(\textbf{r}^{\prime}\right) \phi_{j+k}^{\beta}\left(\textbf{r}\right) \nonumber \\
&& \times V\left(\textbf{r},\textbf{r}^{\prime}\right)\phi_{j}^{\beta}\left(\textbf{r}\right)\phi_{i}^{\alpha}\left(\textbf{r}^{\prime}\right), \label{eqn:vijk}
\end{eqnarray}
where $\alpha,\beta = A,B$.

\subsection{Equations of Motion and Collision Effects}
Using Eq.~\eqref{eqn:mbHamil} and commutation relations for the atomic creation and annihilation operators, we arrive at the following equations of motion for the system:
\begin{eqnarray}
i\hbar\frac{d}{dt}\left<a_i^\dagger a_i\right> & = & -\sum_jg_{ij}\left(\left<b_j^\dagger a_i\right>-\left<a_i^\dagger b_j\right>\right), \\
i\hbar\frac{d}{dt}\left<b_j^\dagger b_j\right> & = & \sum_ig_{ij}\left(\left<b_j^\dagger a_i\right>-\left<a_i^\dagger b_j\right>\right), \\
i\hbar\frac{d}{dt}\left<a_i^\dagger b_j\right> & = & \left[\Delta_{ji}+\sum_{k\neq0}\left(\sigma_{jk}^B\left<b_{j+k}^\dagger b_{j+k}\right>\right.\right. \nonumber \\
&& -\left.\left.\sigma_{ik}^A\left<a_{i+k}^\dagger a_{i+k}\right>\right)\vphantom{\sum_{k\neq0}}\right]\left<a_i^\dagger b_j\right> \nonumber \\
&& -g_{ij}\left(\left<b_j^\dagger b_j\right>-\left<a_i^\dagger a_i\right>\right),
\end{eqnarray}
where $\Delta_{ji} = \epsilon_j^B-\epsilon_i^A$ is the energy shift between the left- and right-hand wells, and $\sigma_{nk}^\alpha = v_{n+k,n,k}^{\alpha\alpha}+v_{n,n+k,-k}^{\alpha\alpha}$, with $\alpha = A,B$, is the coefficient for the energy renormalization.

The final step in developing our model is to include the effects of collisions. Following a treatment that has been successful in describing collisional effects in semiconductor devices, in terms of reproducing the results of quantum kinetic methods~\cite{article:Weng_theory,article:Waldmueller}, we make an effective relaxation rate approximation and write
\begin{equation} \label{eqn:col}
\left.\frac{dn_i^\sigma}{dt}\right|_{\text{col}} = -\gamma\left[n_i^\sigma-f\left(\epsilon_i^\sigma,\mu,T\right)\right],
\end{equation}
where $\gamma$ is an effective collision rate that drives the system towards an equilibrium distribution. Here, $n_i^\sigma$ is the actual population of the $i^{\text{th}}$ level in the $\sigma$ well and $f\left(\epsilon_i^\sigma,\mu,T\right)$ is the equilibrium population that it relaxes to in the limit $\gamma\rightarrow\infty$. In our case, $f\left(\epsilon_i^\sigma,\mu,T\right)$ is the Bose-Einstein distribution function for a given chemical potential, $\mu$, and temperature, $T$, which are determined by number and energy conservation:
\begin{eqnarray}
\sum_{\sigma = A,B}\sum_in_i^\sigma & = & \sum_{\sigma = A,B}\sum_if\left(\epsilon_i^\sigma,\mu,T\right), \label{eqn:nconsv}\\
\sum_{\sigma = A,B}\sum_i\epsilon_i^\sigma n_i^\sigma & = & \sum_{\sigma = A,B}\sum_i\epsilon_i^\sigma f\left(\epsilon_i^\sigma,\mu,T\right). \label{eqn:econsv}
\end{eqnarray}

To picture the collision effects described by Eq.~\eqref{eqn:col}, consider again the scenario of Fig.~\ref{fig:fig1}(a). Assuming the system is completely isolated, the energy of the atoms can only be redistributed by interatomic collisions. Initially, the atomic distribution is far from equilibrium as is evident from the right-hand plot in Fig.~\ref{fig:fig1}(a), where there are empty states in the population distribution of the two-well system at energies corresponding to levels in the right-hand well (e.g., at $E_{R1}$). As the system evolves, collisions drive the system towards equilibrium by redistributing the population among the available states. The end result is an equilibrium population distribution with higher lying states populated (i.e., at a higher effective temperature (see Fig.~\ref{fig:fig1}(b)) because of energy conservation). This is the role of the collisions represented by Eqs.~\eqref{eqn:col}~--~\eqref{eqn:econsv}.

\section{Application of the Model}
To illustrate the application of the above model, we consider what might represent a portion of a device that is used in an atomtronic circuit. It consists of an asymmetric double-well~\cite {article:dw_theory,article:dw_dynamics} in which atom flux arrives at one well and leaves through the other. Of interest is the path by which this simple system arrives at a dynamical steady-state, while also remaining out of thermal equilibrium.

For this example, a further simplification is made by using the following relations
\begin{eqnarray}
\left<a_i^\dagger a_i\right> & = & n_i^A, \\
\left<b_j^\dagger b_j\right> & = & n_j^B, \\
\left<a_i^\dagger b_j\right> & = & \sqrt{n_i^An_j^B}\exp{\left[-i\theta_{ij}\right]},
\end{eqnarray}
where $\theta_{ij} = \theta_i^A-\theta_j^B$ is the relative phase between states of the left- and right-hand wells. Inclusion of collision effects, as well as source and extraction contributions, $P_i$ and $\gamma_i^{A,B}$, respectively, gives
\begin{eqnarray}
\frac{dn_i^A}{dt} & = & -\frac{2}{\hbar}\sum_jg_{ij}\sqrt{n_i^An_j^B}\sin{\theta_{ij}}+\left.\frac{dn_i^A}{dt}\right|_{\text{col}} \nonumber \\ 
&& +P_i-\gamma_i^A n_i^A, \\ \label{eqn:tpopf1}
\frac{dn_j^B}{dt} & = & \frac{2}{\hbar}\sum_ig_{ij}\sqrt{n_i^An_j^B}\sin{\theta_{ij}}+\left.\frac{dn_i^B}{dt}\right|_{\text{col}} \nonumber \\ 
&& -\gamma_j^B n_j^B, \\ \label{eqn:tpopf2}
\frac{d\theta_{ij}}{dt} & = & \frac{1}{\hbar}\left[\Delta_{ji}+\sum_{k\neq0}\left(\sigma_{jk}^Bn_{j+k}^B-\sigma_{ik}^An_{i+k}^A\right)\right] \nonumber \\ 
&& -\frac{g_{ij}}{\hbar}\frac{\left(n_j^B-n_i^A\right)}{\sqrt{n_i^An_j^B}}\cos{\theta_{ij}}, \label{eqn:tpopf3}
\end{eqnarray}

In the simulations we consider the ground state of the left-hand well initially occupied together with a negligible background population in the other states (i.e., initially an essentially zero temperature BEC exists in the left-hand well). We consider six energy levels in the left-hand well and four in the right-hand well.  The initial populations are $n_1^A = 0.08$ and $n_i^A = 10^{-6}$ for $i > 1$ in the left-hand well, and $n_1^B = 0.001$ in the right-hand well.  We use an effective collision rate of $\gamma = 4~\text{s}^{-1}$. Atoms are injected into the fifth level of the left-hand well at a rate of $P_5 = 0.1~\text{s}^{-1}$, and evaporated from the highest levels at a rate of $\gamma_6^A = \gamma_4^B = 4~\text{s}^{-1}$. The population in the lowest three energy levels of the right-hand well are outcoupled at a rate of $\gamma_{1,2,3}^B = 0.1~\text{s}^{-1}$. The temporal evolution of the populations in each well are studied in the presence of collisions and a direct coupling between the ground states of each well as described by Eq.~\eqref{eqn:gij} for $g_{11}$.

In this double-well system the presence of a relative phase between BECs in the left- and right-hand wells is of particular interest. Based on Eq.~\eqref{eqn:tpopf3} phase-locking (i.e. $d\theta_{11}/dt\equiv d\varphi/dt\rightarrow0$) can occur when the term containing $g_{11}$ in Eq.~\eqref{eqn:tpopf3} is larger than the first term on the right-hand side of the same equation. The effect of phase-locking on this system is discussed in more detail below where the time evolution of the system is studied for different coupling strengths between the two wells of the potential.

\subsection{Unlocked operation}
\begin{figure}
\centering
\includegraphics[scale=1]{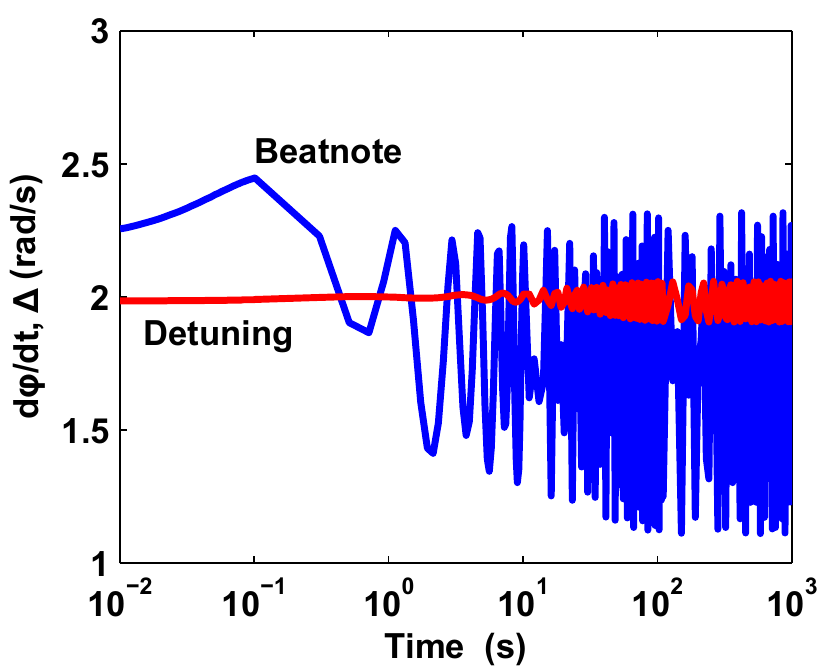}
\caption{\label{fig:fig3} Time evolution of the relative phase (solid, blue) and detuning (solid, red) between ground state populations of the two wells for the unlocked case.}
\end{figure}

We begin by considering a ground state coupling parameter of $g_{11}/\hbar = 1~\text{s}^{-1}$. Figure~\ref{fig:fig3} shows the temporal evolution of $\varphi$.
Also plotted is the ground state frequency difference, or detuning, which is given by 
\begin{equation}
\Delta = \frac{1}{\hbar}\left[\Delta_{11}+\sum_{k\neq0}\left(\sigma_{1,k}^Bn_{1+k}^B-\sigma_{1,k}^An_{1+k}^A\right)\vphantom{\sum_{k\neq0}}\right],
\end{equation}
where $\sigma_{1,k}^{A}/\hbar = \sigma_{1,k}^{B}/\hbar = 0.2~\text{s}^{-1}$ for all $k\neq0$. For $g_{11}/\hbar = 1~\text{s}^{-1}$, the BEC phases are unlocked and the relative phase between the two wells exhibits rapid oscillations. Additionally, the detuning remains nearly constant in time.

\begin{figure}
\centering
\includegraphics[scale=1]{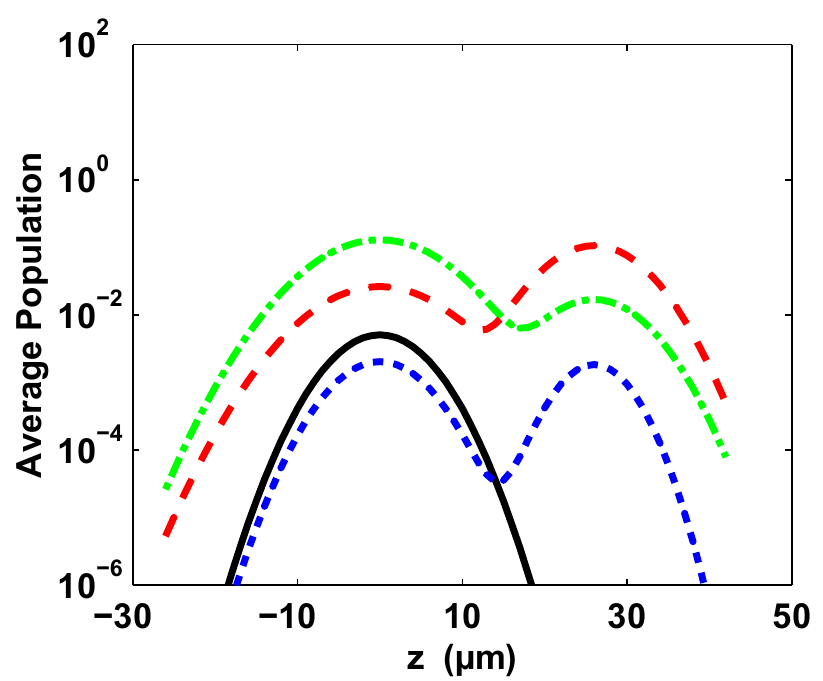}
\caption{\label{fig:fig4} Spatial distribution of the ensemble-averaged atomic population, $\left<\psi^\dagger\psi\right>$, in the unlocked case for $t = 10^{-6}~\text{s}$ (solid, black), $1~\text{s}$ (dotted, blue), $624.5~\text{s}$ (dash-dotted, green), and $626.4~\text{s}$ (dashed, red). At large times the population distribution undergoes small oscillations between the two wells as depicted by the dash-dotted (green) and dashed (red) curves.}
\end{figure}

Figure~\ref{fig:fig4} shows the time evolution of the spatial distribution of the ensemble-averaged atomic population, which is given by
\begin{eqnarray}
\left<\hat{\psi}^\dagger\hat{\psi}\right> & = & \left|\phi_1^A\right|^2n_1^A+\left|\phi_1^B\right|^2n_1^B \nonumber \\
&& +2\sqrt{n_1^An_1^B}\text{Re}\left\{\phi_1^{A}\phi_1^B\text{exp}\left[i\varphi\right]\right\}.
\end{eqnarray}
The initial population distribution is localized in the left-hand well as depicted by the $t = 10^{-6}~\text{s}$ curve. The remaining curves show the temporal evolution of the population distribution where small oscillations between the non-degenerate eigenstates $\phi_1^A$ and $\phi_1^B$ occur, as depicted by the $t = 624.5~\text{s}$ and $626.4~\text{s}$ curves.

\subsection{Locked operation}
\begin{figure}
\centering
\includegraphics[scale=1]{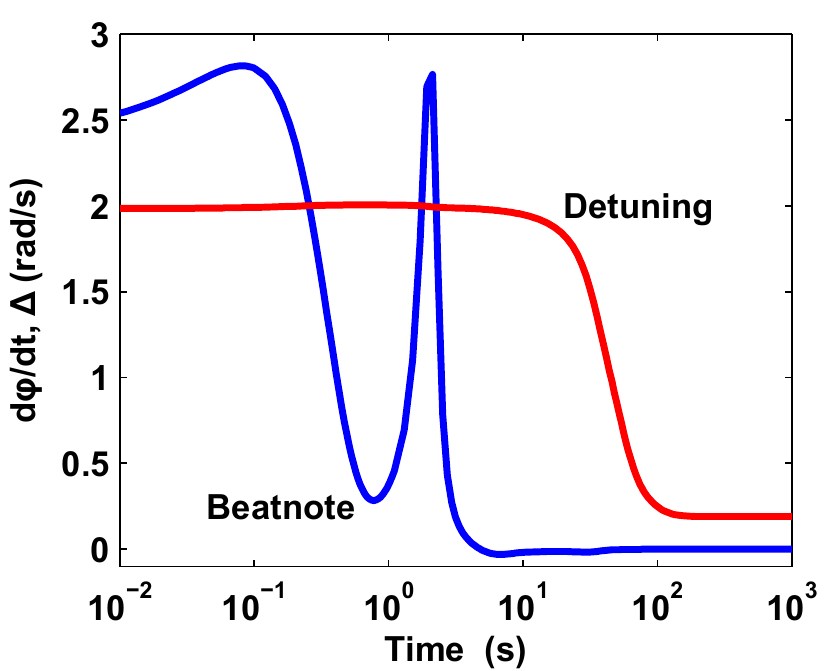}
\caption{\label{fig:fig5} Time evolution of the relative phase (solid, blue) and detuning (solid, red) between ground state populations of the two wells for the locked case.}
\end{figure}

System behavior becomes distinctly different when $g_{11}/\hbar$ increases to $2~\text{s}^{-1}$. Figure~\ref{fig:fig5} explicitly shows the phase-locking, which occurs after a time $t > 2~\text{s}$.
The locking arises from an increase in the ground state population, $n_1^A$, of the left-hand well due to the pump and extraction processes. This in turn causes the exchange shift to decrease the detuning between left- and right-hand well ground states (red curve, Fig.~\ref{fig:fig5}). When the detuning becomes sufficiently small, phase locking occurs (compare blue curves, Figs.~\ref{fig:fig3} and~\ref{fig:fig5}).

\begin{figure}
\centering
\includegraphics[scale=1]{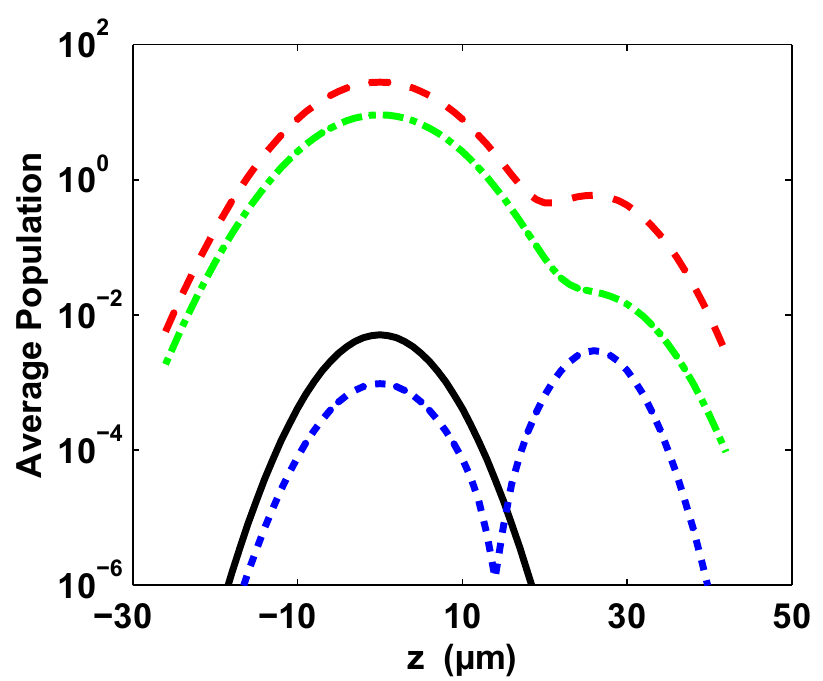}
\caption{\label{fig:fig6} Spatial distribution of the ensemble-averaged atomic population, $\left<\psi^\dagger\psi\right>$, in the locked case for $t = 10^{-6}~\text{s}$ (solid, black), $1~\text{s}$ (dotted, blue), $50~\text{s}$ (dash-dotted, green), and $1000~\text{s}$ (dashed, red). A time-independent distribution is reached where an appreciable population has built up in the left well.}
\end{figure}

Figure~\ref{fig:fig6} shows the evolution of the same initial atomic spatial distribution as in the unlocked case.
Here, the populations in both wells continue to grow during the transfer of atoms from the left-hand to the right-hand well, until a time-independent distribution is reached, with substantially higher populations in both wells than in the unlocked situation (compare to Fig.~\ref{fig:fig4}). Underlying the time-independence and population growth is the formation of a composite eigenstate, resulting from the phase locking of the ground states in both wells.

\subsection{Comparison of locked and unlocked operation}
\begin{figure}
\centering
\includegraphics[scale=1]{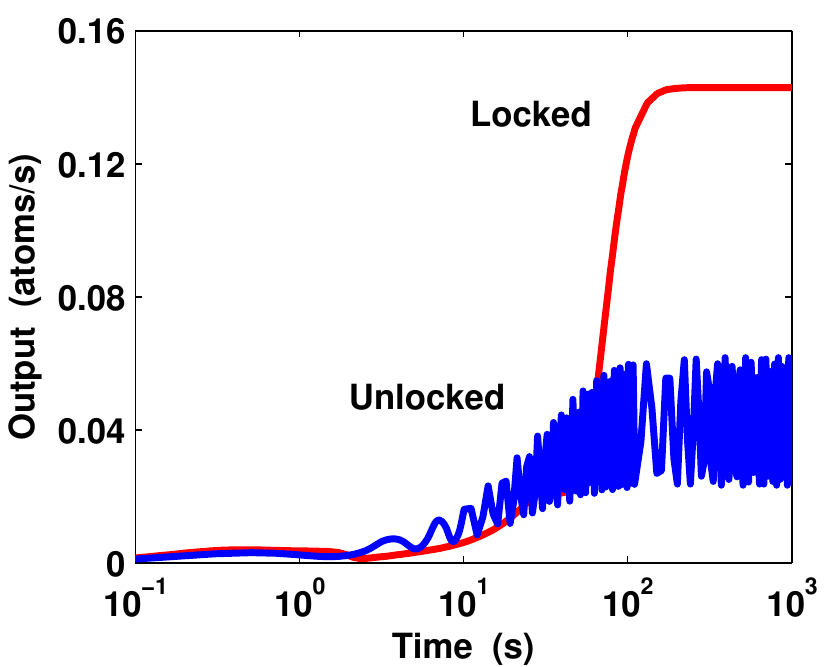}
\caption{\label{fig:fig7}Time evolution of the output flux for the unlocked (solid, blue) and locked (solid, red) case.}
\end{figure}

The significance between locked and unlocked operation for circuit operation is clearly evident when observing the number of atoms leaving the system. In Fig.~\ref{fig:fig7} the atom flux leaving the right-hand well is plotted as a function of time. When the system is not phase-locked the output flux fluctuates around an average value of $\sim0.04~\text{atoms/s}$. The fluctuations arise as a result of a small back and forth population transfer between the two wells. When phase-locking occurs, the output flux increases by a factor of 3.5 to a steady-state value of $\sim0.14~\text{atoms/s}$.

\begin{figure}
\centering
\includegraphics[scale=1]{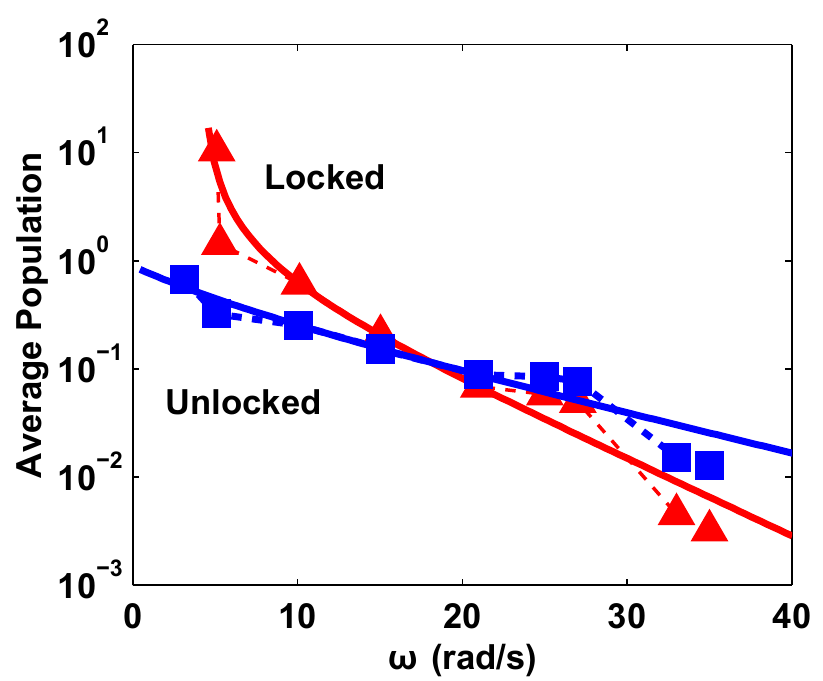}
\caption{\label{fig:fig8} Steady-state population distributions in the unlocked (solid, blue) and locked (solid, red) cases. Each data point corresponds to the population at a given energy level. The solid lines show the Bose-Einstein distributions with equivalent total population and energy.}
\end{figure}

Some insight into the increase in output flux due to phase-locking can be gained by studying the population in each level once steady-state is reached. Figure~\ref{fig:fig8} shows the population distribution in steady-state for both the unlocked and locked cases.
As time progresses in the unlocked case, relaxation takes place and the populations settle essentially to a Bose-Einstein distribution after $t\approx10~\text{s}$. Note that the population distribution is not exactly Bose-Einstein in nature due to the effects of injecting and extracting atoms at specific energy levels. A similar progression is followed at early times in the locked case; however, with the onset of phase-locking substantial cooling takes place (indicated by the decrease in population of higher lying states) while the total atom population continues to grow appreciably. Time independence in this case occurs around $t\approx100~\text{s}$.

\begin{figure}
\centering
\includegraphics[scale=1]{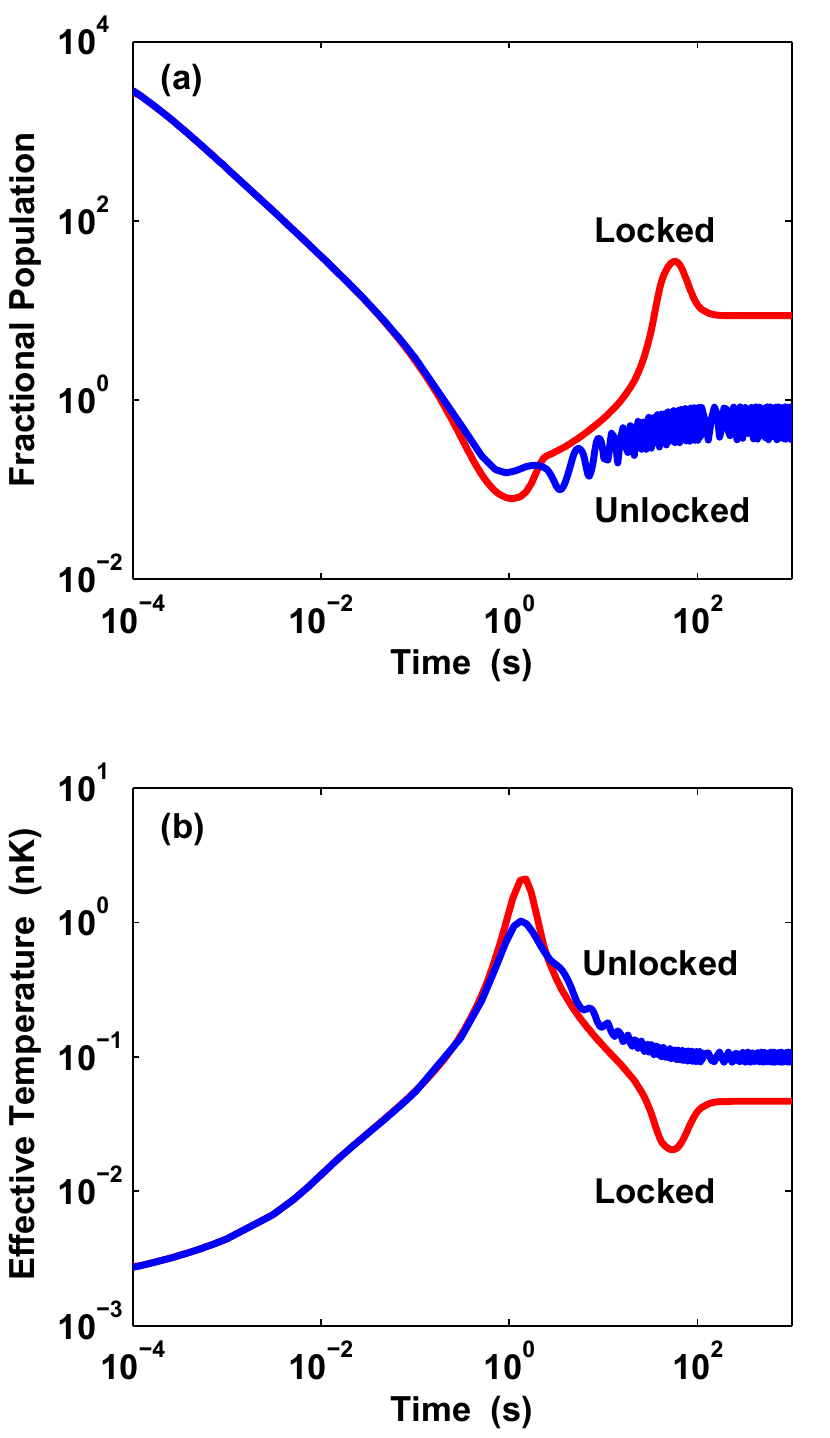}
\caption{\label{fig:fig9} Time evolution of the (a) fractional ground state population and (b) effective temperature of atoms in the left-hand well for the unlocked (solid, blue) and locked (solid, red) cases.}
\end{figure}

Fig.~\ref{fig:fig9} summarizes the results presented in Fig.~\ref{fig:fig8} by showing the time evolution of the fractional ground state population and effective temperature of atoms in the left-hand well.
Here, the fractional ground state population in the left-hand well is given by $n_1^A/\sum_{i=2}^6n_i^A$. This quantity is plotted in Fig.~\ref{fig:fig9}(a), where the difference in condensate occupation is clearly depicted. Figure~\ref{fig:fig9}(b) shows the effective temperature of the left-hand well, which is obtained by fitting the population distribution to a Bose-Einstein distribution. At early times, both locked and unlocked systems show a rise in effective temperature as chemical potential is converted to thermal energy and the system trends towards equilibrium. In the unlocked case the system appears to reach a steady-state with noticeable residual fluctuations. On the other hand, in the locked case the system moves towards a time-independent operation exhibiting significant cooling and BEC growth in the left-hand well. This growth leads to the large increase in the number of atoms flowing out of the right-hand well as shown in Fig.~\ref{fig:fig7}.

The physics of coherent and incoherent tunneling has drawn interest since the $1920$'s. In regards to macroscopic quantum tunneling experiments with BEC, recent reports have also indicated two regimes of operation: the Josephson oscillation regime and the self-trapping regime~\cite{article:Oberthaler}, which are determined by the relative phase of two BECs occupying neighboring potential wells. Direct relation of our modeling results to experimental work is ongoing. The connection is challenging because our model casts system behavior in terms of the energy eigenstates of the wells and how they are effected by population dynamics, such as the exchange shifts. In contrast, the experimental work relates system behavior directly to the atomic populations.

\section{Conclusion}
A model is developed for use in the design of atomtronic circuitry based on finite temperature Bose-condensed gases. Working in the Heisenberg picture, equations of motion for atomic populations and coherences are derived. Energy renormalizations are treated at the mean-field level, and collision effects are taken into account using an effective relaxation rate description. Numerical solution of the population and coherence equations of motion allows tracking of the dynamics of finite temperature BECs in the presence of atom injection and extraction.

The model is demonstrated by studying the evolution of a Bose-condensed gas in the presence of atom injection and extraction in a double-well potential. In the presence of collisions and dissipation, two regimes of device operation are observed corresponding to unlocked and locked operation. These operating regimes are identified based on whether the population transfer between the two wells of the potential is incoherent (unlocked) or coherent (locked). Behavior of the system in terms of the output flux, degree of condensation, and effective temperature is shown to be distinctly different for each operating regime.

The goal of this paper is to introduce a predictive and flexible model for use in designing atomtronic devices and for analyzing BEC experiments. There is much room for further improvement of the model. For example, quantum fluctuations that are especially important at very low temperature may be included by extending the equations of motion to include higher order correlations described by terms like $\left<a_i^\dagger a_i^\dagger a_ia_i\right>$, $\left<b_j^\dagger b_j^\dagger b_jb_j\right>$, and $\left<a_i^\dagger b_j^\dagger b_ja_i\right>$.  In addition, the role of dissipation and dephasing may be treated more consistently by use of the equivalence to the Lindblad terms in the density operator approach~\cite{book:Louisell}.
 
\begin{acknowledgments}
This work is partially supported by the Sandia LDRD program, funded by the U.S. Department of Energy under contract DE-AC04-94AL85000, and by the U.S. Air Force Office of Scientific Research under contract FA9550-14-1-0327, the U.S. National Science Foundation under contract PHY1125844, and the Charles Stark Draper Laboratory under contract SC001-0000000759.
\end{acknowledgments}

\end{document}